\documentclass[12pt]{article}
\usepackage[english,german,french,polish]{babel}
\usepackage[T1]{fontenc}
\usepackage{amsfonts}

\begin{document}

\selectlanguage{english}

\baselineskip 0.77cm
\topmargin -0.6in
\oddsidemargin -0.1in

\let\ni=\noindent

\renewcommand{\thefootnote}{\fnsymbol{footnote}}

\newcommand{\SM}{Standard Model }

\pagestyle {plain}

\setcounter{page}{1}

\pagestyle{empty}

~~~

\begin{flushright}
IFT-- 06/21
\end{flushright}

\vspace{0.4cm}

{\large\centerline{\bf Constraint of parameters in a neutrino mass formula{\footnote{Work supported in part by the Polish Ministry of Science and Higher Education, grant 1 PO3B 099 29 (2005-2007). }}}} 

\vspace{0.4cm}

{\centerline {\sc Wojciech Kr\'{o}likowski}}

\vspace{0.3cm}

{\centerline {\it Institute of Theoretical Physics, Warsaw University}}
 
{\centerline {\it Ho\.{z}a 69,~~PL--00--681 Warszawa, ~Poland}}

\vspace{0.5cm}

{\centerline{\bf Abstract}}

\vspace{0.2cm}

A search for the parameter constraint in the three-parameter empirical mass formula proposed 
recently for active neutrinos is described. Without any parameter constraint the formula is
a formal transformation of three free parameters into three neutrino masses or {\it vice versa}, 
giving no numerical predictions for the masses. But, this is a very special transformation, 
providing some specific suggestions as to its parameters, when it is confronted with the 
present, unfortunately incomplete, experimental data. If the parameters become constrainted, 
then there appear some numerical predictions for the neutrino masses, subject to  
experimental verification (as far as it is or will be possible).

\vspace{0.5cm}

\ni PACS numbers: 12.15.Ff , 14.60.Pq   

\vspace{0.6cm}

\ni October 2006  

\vfill\eject

~~~
\pagestyle {plain}

\setcounter{page}{1}

In a recent paper [1], we have proposed an empirical mass formula for three active mass neutrinos $\nu_1\,,\,\nu_2\,,\,\nu_3$ related to three active flavor neutrinos $\nu_e\,,\,\nu_\mu\,,\,\nu_\tau$ through the unitary mixing transformation $\nu_\alpha =\sum_i U_{\alpha i} \nu_i \;\; (\alpha = e, \mu,\tau \;\;{\rm and}\;\; i = 1,2,3)$. The formula involves three free parameters  $\mu \,,\, \varepsilon \,,\, \xi $ and gets the form

\begin{equation}
m_{i}  = \mu \, \rho_i \left[1- \frac{1}{\xi} \left(N^2_i + \frac{\varepsilon -1}{N^2_i}\right)\right] \;,
\end{equation}

\ni where 

\begin{equation}
N_1 = 1 \;,\; N_2 = 3 \;,\; N_3 = 5 \;\;\;,\;\;\; \rho_1 = \frac{1}{29} \;,\; \rho_2 = \frac{4}{29} \;,\; \rho_3 = \frac{24}{29}
\end{equation}

\vspace{0.1cm}

\ni ($\sum_i \rho_i = 1$). The latter three numbers have been called generation-weighting factors. Since Eq. (1) has the form of a transformation of three free parameters  $\mu \,,\, \varepsilon \,,\, \xi $ into three neutrino masses $m_1\,,\,m_2\,,\,m_3$, the parameters are determined by the masses or {\it vice versa}. In particular, if $m_{1} < m_{2} < m_{3}$ with

\begin{equation}
m_1 \sim 0 \;{\rm to}\;10^{-3}\;{\rm eV} \;,\; \Delta m^2_{21} \sim 8.0\times 10^{-5}\; {\rm eV}^2 \;,\; \Delta m^2_{32} \sim 2.4\times 10^{-3}\; {\rm eV}^2 \;,
\end{equation}

\ni then

\begin{equation}
m_2 \sim (8.9 \;\;{\rm to}\;\; 9.0)\times 10^{-3}\; {\rm eV} \;,\; 
m_3 \sim 5.0\times 10^{-2}\; {\rm eV} \;,
\end{equation}

\vspace{0.2cm}

\ni and we determine 

\begin{equation}
\mu \sim (7.9\;{\rm to} \;7.5)\times 10^{-2}\;{\rm eV} \;,\;\frac{\varepsilon}{\xi} \sim 1\;{\rm to} \;0.61 \;,\; \frac{1}{\xi} \sim (8.1\;{\rm to} \;6.9)\times 10^{-3}\;.
\end{equation}

\ni Of course, there are no numerical predictions for the masses, {\it unless} free parameters are constrained.

In the paper [1], two options of parameter constraint were considered: $\varepsilon/\xi = 1 $  or $1/\xi = 0$. In these cases,

\begin{equation}
m_1 = 0 \;\;{\rm or}\;\; m_3 = \frac{6}{25} (27m_2 - 8 m_1) \;,
\end{equation}

\ni respectively. In the paper [2], the option of parameter constraint $\varepsilon = 0$ was discussed. In this case,

\begin{equation}
m_3 = \frac{6}{125} (351m_2 -  904 m_1) \;. 
\end{equation}

\ni In the present note, we describe the option of parameter constraint $\varepsilon/\xi = 1 - 1/\xi$ that relaxes our first option. Now,

\begin{equation} 
m_3 = \frac{6}{25} (27 m_2 -  1544 m_1) \;. 
\end{equation}

The results of our search in the four options can be listed in the following table:

\vspace{0.2cm}

\begin{center}
\begin{tabular}{c|c|c|c|c|c}
$\varepsilon/\xi$ & $1/\xi(10^{-3})$ & $\mu (10^{-2}\,{\rm eV})$ & $m_1 \,(10^{-3}\,{\rm eV})$ & $m_2\, (10^{-3}\,{\rm eV})$ & $m_3\, (10^{-3}\,{\rm eV})$ \\ 
\hline & & & & & \\
1~~ & 8.1 & 7.9 & 0~~ & 8.9 & 50 \\ -8.8~ & 0~~ & 4.5 & 15~~~ & 12~~~ & 51 \\ 0~~~ & 6.1 & 7.1 & 2.5 & 9.3 & 50 \\ $1 -1/\xi$ & 8.1 & 7.9 & ~~~0.022 & 8.9 & 50    
\end{tabular}
\end{center}

\vspace{0.2cm} 

\ni Here, the experimental estimates $|\Delta m^2_{21}| \sim 8.0\times 10^{-5}\;{\rm eV}^2$ and $\Delta m^2_{32} \sim 2.4\times 10^{-3}\;{\rm eV}^2$ [3] are applied as an input. In the option of
 $1/\xi = 0$, the ordering of $m_1$ and $m_2$ is inverted, though the position of $m_3$ is normal.

In the option of parameter constraint $\varepsilon/\xi = 1 - 1/\xi $, the last in this listing, we can write due to Eq. (8) the following quadratic equation for $ r \equiv m_1/m_2 $:

\begin{equation}
\left[(1544)^2 +\left(\frac{25}{6}\right)^2 \lambda\right] r^2 - 2\cdot 27 \cdot 1544\, r + (27)^2 - \left(\frac{25}{6}\right)^2 (\lambda +1) =0 \,,
\end{equation}

\ni since with $\lambda \equiv \Delta m^2_{32}/\Delta m^2_{21}$ we have

\begin{equation}
m^2_3  \equiv \Delta m^2_{32} + m^2_2 = (\lambda +1) m^2_2 - \lambda\, m^2_1 \,.
\end{equation}

\ni Taking $\lambda \sim 2.4/0.080 = 30$, we obtain two solutions to Eq. (9):

\begin{equation}
r \sim \left\{ \begin{array}{l} 2.46\times 10^{-3} = 2.5 \times 10^{-3}\\ 3.25 \times 10^{-2} = 3.3 \times 10^{-2} \end{array}\right. \,.
\end{equation}

\ni Then, we {\it predict} 

\begin{eqnarray}
m_1 & \equiv & \sqrt {\frac{r^2 \Delta m^2_{21}}{1-r^2}} \sim \left\{ \begin{array}{l} 2.20\times 10^{-5}\;{\rm eV} = 2.2\times 10^{-5}\;{\rm eV} \\ 2.91\times 10^{-4}\;{\rm eV} = 2.9\times 10^{-4}\;{\rm eV} \end{array}\right.\,, \\ 
m_2 & \equiv & \sqrt {\frac{\Delta m^2_{21}}{1-r^2}} \sim \left\{ \begin{array}{l} 8.94\times 10^{-3}\;{\rm eV} = 8.9\times 10^{-3}\;{\rm eV} \\ 8.95\times 10^{-3}\;{\rm eV} = 8.9\times 10^{-3}\;{\rm eV} \end{array}\right.
\end{eqnarray}

\ni and


\begin{equation}
m_3 = \frac{6}{25}\,(27\,m_2 - 1544\,m_1) \sim \left\{\begin{array}{l} 4.98\times 10^{-2}\;{\rm eV} =  5.0\times 10^{-2}\;{\rm eV} \\ -4.98\times 10^{-2}\;{\rm eV} = -5.0\times 10^{-2}\;{\rm eV} \end{array}\right. \,,
\end{equation}

\vspace{0.1cm}

\ni where the experimental estimates $\Delta m^2_{21} \sim 8.0\times 10^{-5}\;{\rm eV}^2$ and  formerly $\Delta m^2_{32} \sim 2.4\times 10^{-3}\;{\rm eV}^2$ in $r$ were used as an imput. So, the second solution (11) for $r$ is unphysical as leading to a negative value of $m_3$ (we assume here that all $m_i$  have the same sign).  Hence, with the first solution (11) for $r$ we determine also

\begin{eqnarray}
\mu = \;\;29 m_1\;\xi \;\; & \sim &  7.93\times 10^{-2}\;{\rm eV} = 7.9\times 10^{-2}\;{\rm eV} \;,\nonumber \\ 
\frac{1}{\xi} = \frac{32 m_1}{9 m_2 + 316 m_1} & \sim & 8.06\times 10^{-3} = 8.1\times 10^{-3} 
\end{eqnarray}

\ni because of the relations

\begin{equation}
m_1 = \frac{\mu}{29}(1 - \frac{\varepsilon}{\xi}) = \frac{\mu}{29}\,\frac{1}{\xi}\;\;,\;\; \frac{1/\xi}{1 - 10/\xi} = \frac{1 - \varepsilon/\xi}{1 - 10/\xi} = \frac{32 m_1}{9 m_2 - 4 m_1}\,.
\end{equation}

\ni They follow in the option of $\varepsilon/\xi = 1 - 1/\xi $ from the neutrino mass formula (1) rewritten explicitly:  

\begin{eqnarray}
m_1 & = & \frac{\mu}{29} (1 - \frac{\varepsilon}{\xi}) \,, \nonumber \\
m_2 & = & \frac{\mu}{29}\, 4\left[1 -\frac{1}{9\xi}(80 + \varepsilon)\right] \,, \nonumber \\
m_3 & = & \frac{\mu}{29} \,24\left[ 1 -\frac{1}{25\xi}\,(624 + \varepsilon)\right] \,.
\end{eqnarray}

\vspace{0.1cm}

In another paper [4], we tried to avoid the input of experimental estimate for $\Delta m^2_{32}$, replacing it by a prediction which follows from the additional conjecture that the off-diagonal part of neutrino mass matrix is built up from annihilation and creation operators acting in the neutrino generation space:

\begin{eqnarray} 
\left( \begin{array}{rrr} 0\:\: & M_{e\,\mu} & M_{e\,\tau} \\ M_{e\,\mu}  & 0\:\: & M_{\mu\,\tau} \\ 
M_{e\,\tau}  & M_{\mu\,\tau} & 0\:\: \end{array}\right) & = & \mu \,\rho^{1/2} \left[g(a + a^\dagger) + g'(a^2 + a^{\dagger\,2})\right] \rho^{1/2} \nonumber \\ & & \nonumber \\ 
& = & \frac{\mu}{29} \left( \begin{array}{ccc} 0 & 2g & \!\!\! 4 \sqrt3 \,g' \\ 2g & 0 & 8\sqrt3 \,g \\ 4 \sqrt3 \,g' & 8\sqrt3 \,g & 0 \end{array}\right)\;\,. 
\end{eqnarray}
 
\vspace{0.2cm}

\ni Here, $g$ and $g'$ are free parameters (multiplied by the mass scale $\mu$ introduced in Eq. (1)) and

\begin{equation} 
\rho^{1/2} = \left( \begin{array}{ccc} \rho^{1/2}_1 & 0 & 0 \\ 0 & \rho^{1/2}_2 & 0 \\ 0 & 0 & \rho^{1/2}_3 \end{array}\right) = \frac{1}{\sqrt{29}} \left( \begin{array}{ccc} 1 & 0 & 0 \\ 0 & \sqrt4 & 0 \\  0 & 0 & \sqrt{24} \end{array}\right)   
\end{equation}
 
\ni according to Eq. (2), while

\begin{equation} 
a = \left( \begin{array}{ccc} 0 & 1 & 0 \\ 0 & 0 & \sqrt2 \\ 0 & 0 & 0 \end{array}\right)  \; ,\;a^\dagger = \left( \begin{array}{ccc} 0 & 0 & 0 \\ 1 & 0 & 0 \\ 0 & \sqrt2 & 0 \end{array}\right)  
\end{equation}

\ni play the role of annihilation and creation operators in the generation space{\footnote{Formally, $ M = (M_{\alpha\, \beta})$ and $ a = (a_{i\,j})$, so we assume that $\rho^{1/2}  = (\delta_{\alpha\, i}\rho_i^{1/2})$, where $\delta_{\alpha\, i} =1$ for ${\alpha\, i} = e\,1,\mu\,2,\tau\,3,$ and 0 otherwise ($\alpha,\beta = e, \mu, \tau$ and $i, j = 1,2,3$). }}, since

\vspace{0.2cm}

\begin{equation} 
n \equiv a^\dagger\, a = \left( \begin{array}{ccc} 0 & 0 & 0 \\ 0 & 1 & 0 \\ 0 & 0 & 2  \end{array}\right)  \;\;,\;\; [a\,,\, n] = 0 \;\;,\;\; [a^\dagger\,,\, n] = -a^\dagger \;\;,\;\; a^3 = 0 \;\;,\;\;
a^{\dagger\,3} = 0 \;,
\end{equation}

\vspace{0.1cm}

\ni though $[ a\,,\, a^\dagger] \neq $ {\bf 1}. Note that the numbers $N_i = 1,3,5$, introduced in this paper in Eqs. (2) and appearing in the mass formula (1), are eigenvalues of the matrix $N \equiv 2n + {\bf 1}$, as the formal occupation-number matrix $n$ has the eigenvalues $n_i = 0,1,2$ corresponding to the generations $i = 1,2,3$ (see Appendix).

It followed immediately from the conjecture (18) that [4]


\begin{equation} 
M_{\mu\,\tau} = 4\sqrt3 \,M_{e\,\mu} \;.
\end{equation}

\vspace{0.2cm}

\ni Making use of the tribimaximal form of neutrino mixing [5] as a reasonable approximation,

\begin{equation} 
U = (U_{\alpha\,i}) = \left( \begin{array}{rrr} \frac{\sqrt2}{\sqrt3} & \frac{1}{\sqrt3} & 0\, \\ - \frac{1}{\sqrt6} & \frac{1}{\sqrt3} & \frac{1}{\sqrt2} \\ \frac{1}{\sqrt6} & -\frac{1}{\sqrt3} & \frac{1}{\sqrt2}  \end{array} \right) \,,
\end{equation}

\ni which implies the neutrino mass matrix $ M = (M_{\alpha\, \beta}) = \left(\sum_i U_{\alpha\,i} \,m_i\, U_{\beta\,i}^* \right)$ with elements

\begin{eqnarray}
M_{e\,e} & = & \:\:\:\frac{1}{3}(2m_1 + m_2)\,, \nonumber \\
M_{\mu\,\mu} & = & \:\:\,M_{\tau\,\tau} = \:\:\:\,\frac{1}{6} (m_1 + 2m_2 + 3 m_3)\,, \nonumber \\
M_{e\,\mu} & = & -M_{e\,\tau} =  -\frac{1}{3} (m_1 - m_2)\,, \nonumber \\
M_{\mu\,\tau} & = & -\frac{1}{6} (m_1 + 2m_2 - 3 m_3)\,,
\end{eqnarray}

\ni we deduced from the relation (22) between $M_{e\,\mu}$ and $M_{\mu\,\tau}$ the following mass sum rule [4]:

\begin{equation}
m_3  =  \eta \,m_2 - (\eta - 1) \,m_1 \,,
\end{equation}
 
\ni where
 
\begin{equation}
\eta \equiv \frac{2}{3} \,(4\sqrt3 +1) = 5.28547\,.
\end{equation}

\ni From the conjecture (18) we also concluded that

\begin{equation} 
g  = - 2\sqrt3 \,g' = \frac{29}{6\mu} \,(m_2 - m_1) \;.
\end{equation}

\ni The mass sum rule (25) together with the mass formula (17) imposed the following relation on $\varepsilon/\xi$ and $1/\xi$:

\begin{equation} 
\frac{1}{\xi} = 0.0173763 - 0.0070452 \frac{\varepsilon}{\xi} \,.
\end{equation}

In the option of parameter constraint $\varepsilon/\xi = 1$ considered in the paper [4], we {\it predicted} with the use of mass sum rule (25) that

\begin{equation}
m_1 = 0 \;,\;   m_2 \sim 8.9\times 10^{-3}\;{\rm eV} \;,\; m_3 \sim 4.7\times 10^{-2}\;{\rm eV} \,.
\end{equation}

\ni Here, {\it only} the experimental estimate $\Delta m^2_{21} \sim 8.0\times 10^{-5}\;{\rm eV}^2$ was applied as an input. We determined also

\begin{equation}
g = - 2\sqrt{3} \,g' \sim 0.53 
\end{equation}

\ni and

\begin{equation}
\mu \sim 8.1\times 10^{-2} \;{\rm eV}\;,\; \frac{1}{\xi} = 1.03311\times 10^{-2} \;,
\end{equation}

\ni the last value consistent with the relation (28). In this case, 
 
\begin{equation}
\lambda \equiv \frac{\Delta m^2_{32}}{\Delta m^2_{21}} = \eta^2 - 1 = 26.9362 \,\end{equation}

\ni and

\begin{equation}
\Delta m^2_{32} \sim 2.15\times 10^{-3}\; {\rm eV}^2 = 2.2\times 10^{-3}\; {\rm eV}^2 \,,
\end{equation}
 
\ni while the popular experimental best fit is $\Delta m^2_{32} \sim 2.4\times 10^{-3}\; {\rm eV}^2$.

In the present note, we consider the relaxation of parameter constraint $\varepsilon/\xi = 1$ to the form $\varepsilon/\xi = 1 - 1/\xi $. Then, the new constraint together with the relation (28) between $\varepsilon/\xi $ and $1/\xi $, valid in the case of mass sum rule (25), lead to the parameter values
 
\begin{equation}
\frac{\varepsilon}{\xi} = 0.989596 \;,\; \frac{1}{\xi} = 0.0104044  \,.
\end{equation}

\ni Then, the mass formula (17), taking in the case of parameter constraint $\varepsilon/\xi = 1 - 1/\xi $ the form

\begin{eqnarray}
m_1 & = & \frac{\mu}{29}\, \frac{1}{\xi} \,, \nonumber \\
m_2 & = & \frac{\mu}{29}\, \frac{4}{9} \left(8 - \frac{79}{\xi}\right) \,, \nonumber \\
m_3 & = & \frac{\mu}{29} \, \frac{24}{25} \left(24 - \frac{623}{\xi}\right) \,,
\end{eqnarray}
 
\ni implies
 
\begin{equation}
r \equiv \frac{m_1}{m_2} = \frac{9}{4(8\xi - 79)} = 3.26\times 10^{-3} = 3.3\times 10^{-3}\,.
\end{equation}

\ni Thus, with the experimental estimate $\Delta m^2_{21} \sim 8.0\times 10^{-5}\; {\rm eV}^2$ as an input, we {\it predict}
 
\begin{eqnarray}
m_1 & \equiv & \sqrt {\frac{r^2 \Delta m^2_{21}}{1-r^2}} \sim 2.92\times 10^{-5}\;{\rm eV} = 2.9\times 10^{-5}\;{\rm eV} \,, \\ 
m_2 & \equiv & \sqrt {\frac{\Delta m^2_{21}}{1-r^2}} \sim  8.94\times 10^{-3}\;{\rm eV} = 8.9\times 10^{-3}\;{\rm eV} 
\end{eqnarray}

\ni and


\begin{equation}
m_3 \equiv \eta\, m_2 -(\eta - 1) m_1 \sim 4.71\times 10^{-2}\;{\rm eV} =  4.7\times 10^{-2}\;{\rm eV} \,.
\end{equation}

\ni We determine also

\begin{equation}
g = - 2\sqrt{3} \,g' = 0.530 = 0.53
\end{equation}

\ni and

\begin{equation}
\mu = 29\, m_1 \xi \sim 8.13\times 10^{-2} \;{\rm eV} = 8.1\times 10^{-2} \;{\rm eV}\;. 
\end{equation}

\ni In this case,

\begin{equation}
\lambda \equiv \frac{\Delta m^2_{32}}{\Delta m^2_{21}} \sim 26.8 = 27 \;\left(< \eta^2 - 1 =26.9362 \right)
\end{equation}

\ni and

\begin{equation}
\Delta m^2_{32} \sim 2.14\times 10^{-3}\; {\rm eV}^2 = 2.1\times 10^{-3}\; {\rm eV}^2 \,.
\end{equation}
 
\ni The values (38), (39) and (41) do not differ practically from the values (29) and (31) for $m_2$, $m_3$ and $\mu $ implied in the option of $\varepsilon/\xi = 1$.

In the case of mass sum rule (25), the results of our search in two options can be compared in the following listing: 

\vspace{0.2cm}

\begin{center}
\begin{tabular}{c|c|c|l|c|c}
$\varepsilon/\xi$ & $1/\xi(10^{-3})$ & $\mu (10^{-2}\,{\rm eV})$ & $m_1 \,(10^{-3}\,{\rm eV})$ & $m_2\, (10^{-3}\,{\rm eV})$ & $m_3\, (10^{-3}\,{\rm eV})$ \\ 
\hline & & & & & \\
1~~ & 10.3311 & 8.1 & ~~~0 & 8.9 & 47 \\ $1 -1/\xi$ & 10.4044 & 8.1 & ~~~0.029  & 8.9 & 47     
\end{tabular}
\end{center}

\vspace{0.2cm} 

\ni Here, {\it only} the experimental estimate $\Delta m^2_{21} \sim 8.0\times 10^{-5}\; {\rm eV}^2$ is used as an input. The {\it prediction} is $\Delta m^2_{32} \sim 2.2\times 10^{-3}\; {\rm eV}^2$ and $2.1\times 10^{-3}\; {\rm eV}^2$, respectively, while the popular experimental best fit is $\Delta m^2_{32} \sim 2.4\times 10^{-3}\; {\rm eV}^2$.

Concluding, we can see from both our listings that in the options of parameter constraint $\varepsilon/\xi = 1$ and $\varepsilon/\xi = 1 - 1/\xi$ the results are rather similar, though in the first of them the lowest mass $m_1$ takes its smallest possible value 0 (in both listings). The option of parameter constraint $\varepsilon/\xi = 0$ differs from the two previous by a much larger value of $m_1$ that now is not dramatically smaller than the value of $m_2$. Eventually, in the option of parameter constraint $1/\xi = 0$, the ordering of $m_1$ and $m_2$ is inverted, though the position of $m_3$ is normal. Thus, if the value of $m_1$ turned out to be not dramatically smaller than the value of $m_2$, the option of $\varepsilon/\xi = 0$ would be favored (among the possibilities considered in this paper).  This option, discussed in Ref. [2], seems to be attractive from the interpretative point of view taking into account a formal "intrinsic structure"\, of neutrinos of three generations (see Section 4 of Ref. [2]; there our present $\mu$, $\varepsilon$, $\xi$ are denoted as $\mu^{(\nu)}_{\rm eff }$, $\varepsilon^{(\nu)}$, $\xi^{(\nu)}$; see also Appendix). From this point of view, the alternative option of $\varepsilon/\xi = 1 - 1/\xi \stackrel{<}{\sim} 1$ described in the present paper (in both our listings) seems also attractive. But in this case, $m_1 = (\mu/29)(1 - \varepsilon/\xi) = (\mu/29)/\xi$ would be very small $ \sim O(10^{-5}\;{\rm eV}) \ll m_2 < m_3$. In the case of $\varepsilon = 0$, the whole mass $m_1$ comes out from the formal "intrinsic binding energy"\, of $\nu_1$ {\it via} a simple version of seesaw mechanism [2], while in the case of  $\varepsilon = \xi - 1\stackrel{<}{\sim} \xi $ the formal "intrinsic binding energy"\, of $\nu_1$ would only slightly prevail over its formal "intrinsic selfenergy", their tiny difference providing the very small mass $m_1${\it via} the simple version of seesaw mechanism.

\vspace{0.3cm}

\vfill\eject

{\centerline{\bf Appendix}} 

\vspace{0.3cm}

{\centerline{\it More about the "intrinsic structure"\, of fundamental fermions}} 

\vspace{0.3cm}

Consider the generation triplet of fields describing the fundamental fermions $f_i = \nu_i, l_i, u_i, d_i \;(i = 1,2,3)$ of four kinds $f = \nu, l, u,d $ (leptons and quarks belonging to three generations):


$$
\psi^{(f)}(x) = \left(\begin{array}{l} \psi^{(f_1)}(x) \\ \psi^{(f_2)}(x) \\ \psi^{(f_3)}(x) \end{array}\right) \;.
\eqno{\rm (A1)}
$$


\ni We can also write

$$
\psi^{(f)}(x) = | 0 >\,\psi^{(f_1)}(x) + | 1 >\,\psi^{(f_2)}(x) + | 2 >\,\psi^{(f_3)}(x) \;,
\eqno{\rm (A2)}
$$

\ni where

$$
| 0 > = \left(\begin{array}{c} 1 \\ 0 \\ 0 \end{array}\right) \;,\; 
| 1 > = \left(\begin{array}{c} 0 \\ 1 \\ 0 \end{array}\right) \;,\; 
| 2> = \left(\begin{array}{c} 0 \\ 0 \\ 1 \end{array}\right) \;,
\eqno{\rm (A3)}
$$

\ni and so $ \psi^{(f_i)}(x) = <n_i| \psi^{(f)}(x)>$ with $n_i = 0,1,2 \;(i=1,2,3)$. Then, the matrices (20), $a^\dagger $ and $a$, play the role of universal creation and annihilation operators in generation space, because

$$
a^\dagger |0> = |1> \;,\; a^\dagger |1> = \sqrt{2}|2> \;,\; a^\dagger |2> = 0 
\eqno{\rm (A4)}
$$

\ni and

$$
a |2> = \sqrt2|1> \;,\; a |1> = |0> \;,\; a |0> = 0  \,.
\eqno{\rm (A5)}
$$

\ni Here, the label $n_i \;(i = 1,2,3)$ in basic kets (A3) is given by the eigenvalues $n_i = 0,1,2$ of the formal occupation-number operator $n \equiv a^\dagger a$. The occupation numbers $n_i = 0,1,2$ , or the related numbers $N_i \equiv 2n_i + 1 = 1,3,5$ being eigenvalues of the operator $N \equiv 2n + {\bf 1}$, may label three fundamental-fermion generations $i = 1,2,3$. The conditions $a^3 = 0$ and $a^{\dagger\,3} = 0$ in Eqs. (21) restrict the integers $n_i$ to the range $0 \leq n_i \leq 2$ and so, the number of generations to three.

In our previous work [6], we have proposed an interpretation of the numbers $n_i$ and $N_i \;(i = 1,2,3)$ on the base of a new generalized Dirac equation following from the general Dirac 
square-root procedure. This equation leads to multicomponent wave functions or fields $\psi_{\alpha_1\,\alpha_2 \ldots \alpha_N)}(x) \;(N = 1,2,3,\ldots)$, where $\alpha_k = 1,2,3,4 \;(k = 1,2,...,N)$ are Dirac bispinor indices. We have conjectured that all Dirac bispinor indices $\alpha_k$ but one denoted by $\alpha_1$ obey the (intrinsic) Fermi statistics along with the (intrinsic) Pauli principle ({\it i.e.}, $\alpha_2, \ldots, \alpha_N $ are fully antisymmetrized). Then, it follows that the number of all bispinor indices $\alpha_k \; (k = 1,2,\ldots ,N)$ cannot exceed the maximal number $N =5$. This implies that for fundamental fermions the number $N$ can be equal only to 1 or 3 or 5, what justifies the existence in Nature of {\it exactly} three generations of fundamental fermions. These may be labelled by the numbers $N_i  = 1,3,5$ or $n_i \equiv (N_i - 1)/2 =0,1,2$ corresponding to $i = 1,2,3$, respectively. We can see that $n_i$ are the numbers of pairs of the fully antisymmetrized Dirac bispinor indices, while $N_i$ are the numbers of all Dirac bispinor indices.

In consequence of the above construction, the fundamental fermions $f_i = \nu_i, l_i, u_i, d_i$ of three generations $i = 1,2,3$ can be described by the following three wave functions or fields built up from $\psi^{(f)}_{\alpha_1\alpha_2 \ldots \alpha_{N_i}}(x) \;(N_i = 1,3,5)$:  

\begin{eqnarray*}
\psi^{(f_1)}_{\alpha_1}(x) & \equiv & \psi^{(f)}_{\alpha_1}(x) \;, \\
\psi^{(f_2)}_{\alpha_1}(x) & \equiv & \frac{1}{4}\left( C^{-1} \gamma^5\right)_{\alpha_2\,\alpha_3}\psi^{(f)}_{\alpha_1\,\alpha_2\,\alpha_3}(x) = \psi^{(f)}_{\alpha_1\,1\,2}(x) =\psi^{(f)}_{\alpha_1\,3\,4}(x)  \;, \\
\psi^{(f_3)}_{\alpha_1}(x) & \equiv & \frac{1}{24} \varepsilon_{\alpha_2\alpha_3\alpha_4\alpha_5}\, \psi^{(f)}_{\alpha_1\alpha_2\alpha_3\alpha_4\alpha_5}(x)  = \psi^{(f)}_{\alpha_1\,1234}(x)\;. \\
\end{eqnarray*}

\vspace{-2.46cm}

\begin{flushright}
({\rm A}6)
\end{flushright}

\vspace{-0.3cm}

\ni It can be seen that (due to the full antisymmetry of $\alpha_2,\ldots, \alpha_{N_i}$ indices) the multicomponent wave functions or fields corresponding to $N_i = 1,3,5$ appear (up to the sign $\pm$) with the multiplicities

$$
1 = 29 \rho_1 \;,\; 4 = 29 \rho_2 \;,\; 24 =  29 \rho_3 
\eqno{\rm (A7)}
$$

\ni ($\sum_i \rho_i = 1$), respectively, where $ \rho_i$ are the generation-weighting factors introduced in this paper in Eq. (2) and appearing in the mass formula (1). Hence, the bilinear relations

$$
\psi^{(f)*}_{\alpha_1 \alpha_2 \ldots \alpha_{N_i}}\!(x)\, \psi^{(f)}_{\alpha_1 \alpha_2 \ldots \alpha_{N_i}}\!(x)  = 29 \,\rho_i\,  \psi^{(f_i)*}_{\alpha_1}(x)\, \psi^{(f_i)}_{\alpha_1}(x) \\
\eqno{\rm (A8)}
$$

\ni hold for $N_i = 1,3,5$ (or $i=1,2,3$). 

The numbers $N_i$ (the total numbers of Dirac bispinor indices) and $\rho_i$ (the generation-weighting factors) are at our disposal to describe at our level a formal "intrinsic structure"\, of fundamental fermions [6], in particular, to construct their empirical mass formula [1] (note that these fermions are still pointlike, due to the generalized Dirac equation which is a local differential equation in spacetime).

In the paper [1], it has been proposed that the following empirical mass formula holds for fundamental fermions $f_i = \nu_i, l_i, u_i, d_i \;(i = 1,2,3)$:

$$
m_{f_i}  =  \mu^{(f)}\, \rho_i \left(N^2_i + \frac{\varepsilon^{(f)} -1}{N^2_i} - \xi^{(f)} \right) \;,
\eqno{\rm (A9)}
$$

\ni where $\mu^{(f)}$, $\varepsilon^{(f)}$, $\xi^{(f)}$ are free parameters. The correctness of this mass formula has been successfully tested in the case of charged leptons ($f = l$) whose mass spectrum is known precisely [6]. Then, $\xi^{(l)} \simeq 0$.

In the case of active neutrinos ($f = \nu$), it has been conjectured that this formula works for the neutrino Dirac masses $m_{\nu_i}^{(D)}$, while the neutrino effective masses $m_{\nu_i}$ are induced by a simple version of the seesaw mechanism, where  $m_{\nu_i} = -m_{\nu_i}^{(D)\,2}/M_{\nu_i} = -m^{(D)}_{\nu_i}/\zeta $ with large neutrino Majorana masses $M_{\nu_i}$ assumed to be proportional to  $m_{\nu_i}^{(D)}$: $\;M_{\nu_i} = \zeta m_{\nu_i}^{(D)}$ with a large parameter $\zeta >0$. Then [1],

$$
m_{\nu_i} = \mu^{(\nu)}_i  \rho_i \left[1-  \frac{1}{\xi^{(\nu)}} \left(N^2_i + \frac{\varepsilon^{(\nu)} -1}{N^2_i}\right)\right] \;,
\eqno{\rm (A10)}
$$

\ni where $\mu^{(\nu)}_{\rm eff} \equiv \mu^{(\nu)} \xi/\zeta$, $\varepsilon^{(\nu) }/\xi^{(\nu)}$, $1/\xi^{(\nu)}$ are free parameters. With $m_{\nu_i} \rightarrow m_i$ and $\mu^{(\nu)}_{\rm eff} \rightarrow \mu$, $\varepsilon^{(\nu)} \rightarrow \varepsilon $, $\xi^{(\nu)} \rightarrow \xi$ this equation is the neutrino mass formula (1) discussed in the present paper.

\vfill\eject

~~~~
\vspace{0.5cm}

{\centerline{\bf References}}

\vspace{0.5cm}

{\everypar={\hangindent=0.6truecm}
\parindent=0pt\frenchspacing

{\everypar={\hangindent=0.6truecm}
\parindent=0pt\frenchspacing

\vspace{0.2cm}

[1]~W. Kr\'{o}likowski, {\it Acta Phys. Pol.} {\bf B 37}, 2601 (2006) [{\tt hep--ph/0602018}]; {\it cf.} also {\tt hep--ph/0604148}.

\vspace{0.2cm}

[2]~W. Kr\'{o}likowski, {\tt hep--ph/0609187}.

\vspace{0.2cm}

[3]~{\it Cf. e.g.} G.L. Fogli, E. Lisi, A. Marrone, A. Palazzo, {\it Progr. Part. Nucl. Phys.} {\bf 57}, 742 (2006) [{\tt hep--ph/0506083}]; G.L. Fogli, E. Lisi, A. Mirizzi, D.~Montanino, P.D.~Serpico, 
\tt hep--ph/0608321}.

\vspace{0.2cm}

[4]~W. Kr\'{o}likowski, {\it Acta Phys. Pol.} {\bf B 37}, 2805 (2006) [{\tt hep--ph/0606223}]; 
{\tt hep--ph/0608229}.

\vspace{0.2cm}

[5]~L. Wolfenstein,  {\it Phys. Rev.} {\bf D 18}, 958 (1978); P.F. Harrison, D.H. Perkins, W.G.~Scott, {\it Phys. Lett.} {\bf B 458}, 79 (1999); {\it Phys. Lett.} {\bf B 530}, 167 (2002); Z.Z.~Xing. {\it Phys. Lett.} {\bf B 533}, 85 (2002); P.F. Harrison, W.G.~Scott, {\it Phys. Lett.} {\bf B 535}, 163 (2003); T.D.~Lee, {\tt hep--ph/0605017}.

\vspace{0.2cm}

[6]~W. Kr\'{o}likowski, {\tt hep--ph/0604148}; {\it cf.} also {\it Acta Phys. Pol.} {\bf B 33}, 2559 (2002) [{\tt hep--ph/0203107}]; {\tt hep-ph/0504256}; and references therein.

\vfill\eject

\end{document}